\documentclass[english]{IEEEtran}
\usepackage[T1]{fontenc}
\usepackage[latin9]{inputenc}
\usepackage{graphicx}

\makeatletter

\newcommand{\mathcircumflex}[0]{\mbox{\^{}}}

\providecommand{\tabularnewline}{\\}

\makeatother

\usepackage{babel}
\begin{document}
\title{ChatGPT \emph{vs.} Lightweight Security: First Work Implementing the
NIST Cryptographic Standard ASCON}
\author{Alvaro Cintas-Canto, Jasmin Kaur, Mehran Mozaffari-Kermani, and Reza
Azarderakhsh}
\IEEEspecialpapernotice{\thanks{A. Cintas-Canto is with the School of Technology and Innovation, Marymount
University, Virginia, VA 22207, USA (e-mail: acintas@marymount.edu).}\thanks{J. Kaur and M. Mozaffari-Kermani are with the Department of Computer
Science and Engineering, University of South Florida, Tampa, FL 33620,
USA (e-mail: jasmink2@usf.edu, mehran2@usf.edu).}\thanks{R. Azarderakhsh is with the Department of Computer and Electrical
Engineering and Computer Science, Florida Atlantic University, Boca
Raton, FL 33431, USA (e-mail: razarderakhsh@fau.edu).} }
\maketitle
\begin{abstract}
This study, to the best of our knowledge, is the first to explore
the intersection between lightweight cryptography (LWC) and advanced
artificial intelligence (AI) language models. LWC, in particular the
ASCON algorithm which has been selected as the LWC standard by the
National Institute of Standards and Technology (NIST) in Feb. 2023,
has become increasingly significant for preserving data security given
the quick expansion and resource limitations of Internet of Things
(IoT) devices. On the other hand, OpenAI's large language model (LLM)
ChatGPT has demonstrated significant potential in producing complex,
human-like text. This paper offers a novel method for implementing
the NIST LWC standard, ASCON, using the GPT-4 model. Moreover, this
paper details the design and functionality of ASCON, the procedures
and actual Python implementation of ASCON using ChatGPT, and a discussion
of the outcomes. The results contribute valuable insights into the
efficient application of advanced AI language models in cryptography,
particularly in constrained environments. Source code can be found
at: https://github.com/DrCintas/ASCON-with-ChatGPT.
\end{abstract}


\section{Introduction}

Lightweight cryptography (LWC) has emerged as an approach to secure
resource-constrained devices such as Internet of Things (IoT) devices,
a category that has seen exponential growth over the last few years.
IoT devices, given their severe resource constraints, demand cryptographic
solutions that are capable of working efficiently with such limitations.
This includes limited processing power, small memory sizes, constrained
energy supply, and low-bandwidth communication. Despite these, IoT
devices are expected to maintain the confidentiality, integrity, and
availability of data. Therefore, LWC is designed to provide robust
security for such limited-resource devices. Due to the need of adopting
a LWC standard, the National Institute of Standards and Technology
(NIST) recently concluded its multi-year effort to find it, where
ASCON {[}1{]} emerged as the winner among the top 10 NIST LWC finalists
{[}2{]} in Feb. 2023 {[}3{]}. ASCON is an authenticated encryption
algorithm and associated data scheme which is designed to provide
strong security, high efficiency, and simplicity. Its reduced gate
count and its proven security make it an ideal solution for IoT devices.

With the release of ChatGPT, a large language model (LLM) announced
by OpenAI, the field of deep learning and language models has been
revolutionized. Its implications in academic writing, search engines,
and coding are significant, with particularly impressive results shown
when designing computer systems to satisfy particular user requirements
{[}4{]}. As of now, ChatGPT has two different deep learning-based
language models: GPT-3.5 and GPT-4. The term ``GPT'' stands for
generative pre-trained transformer, a powerful artificial intelligence
language model that uses deep learning to generate human-like text
based on provided input. The GPT-4 model stands out for its neural
network design, which has a substantially higher number of parameters
than previous versions. The neural network employed by the GPT-4 version
comprises significantly more parameters than other previous versions.
This increase in parameter volume is essential for improving the system's
capacity to produce sophisticated and contextually rich text, making
it more useful and efficient even in contexts with limited resources.

However, it is yet to be seen how well ChatGPT performs in writing
complex codes required for cryptographic algorithms. Kwon et. al.
{[}5{]} devise methodologies to implement the current symmetric-key
cryptography block ciphers Advanced Encryption Standard (AES) and
CHAM using ChatGPT; the implementation was shown to be straightforward
and precise, where the generated code compiled without errors. This
paper, for the first time, devises an approach to efficiently utilize
the sophisticated OpenAI LLM ChatGPT to implement the NIST LWC standard
ASCON. Throughout this paper, the model version used to implement
ASCON is GPT-4.

\section{Preliminaries}

\begin{figure*}
\begin{centering}
\includegraphics[scale=0.9]{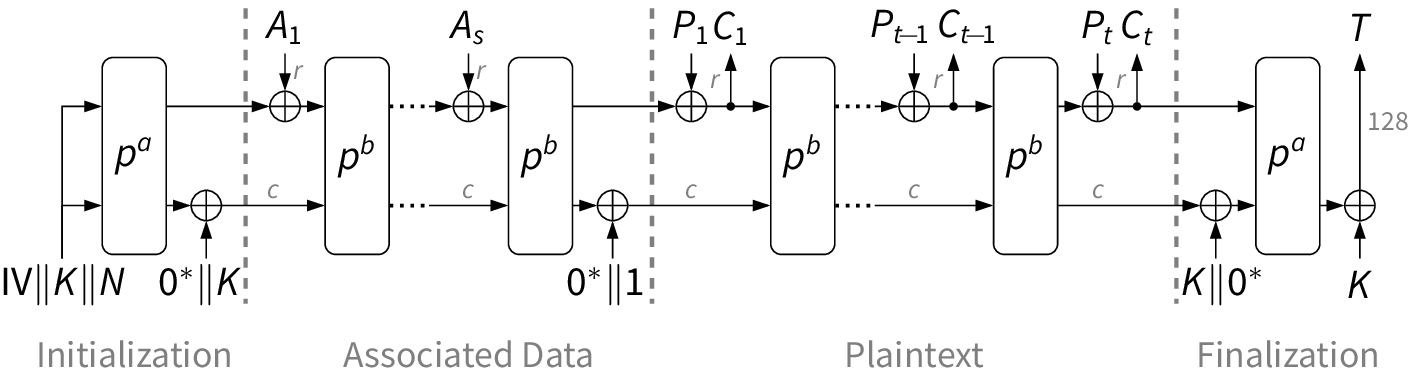}
\par\end{centering}
\caption{ASCON's authenticated encryption process with associated data.}
\end{figure*}
\begin{figure}
\begin{centering}
\includegraphics[scale=0.4]{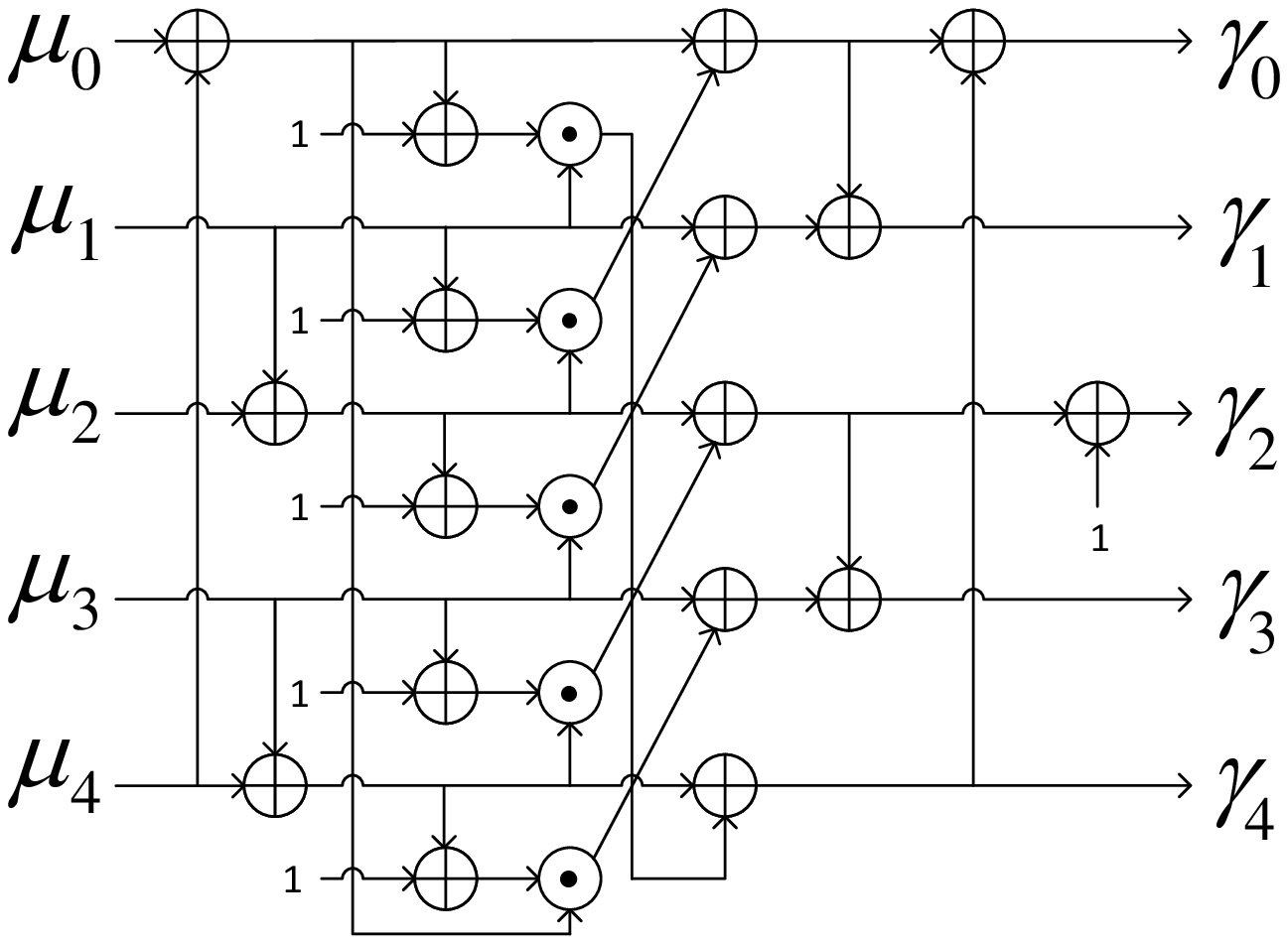}
\par\end{centering}
\caption{Circuit of ASCON's 5-bit S-box.}
\end{figure}
ASCON is an LWC cipher suite with a collection of authenticated encryption
designs and a family of hash and extensible output functions. However,
throughout this work, we are interested in authenticated encryption.
Therefore, for the sake of brevity, we only discuss ASCON authenticated
encryption. For those readers interested in more details about ASCON,
please refer to {[}1{]}. ASCON uses a duplex-based mode of operation,
where ASCON's 320-bit permutation iteratively applies a substitution-permutation
network (SPN) to encrypt or decrypt data. The process takes place
in a bit-slice fashion, a feature that makes it scalable to 8-, 16-,
32-, and 64-bit platforms while remaining lightweight. 
\begin{table}
\centering{}\caption{Parameter sets for the ASCON variants in bits}
\begin{tabular}{|c|c|c|c|c|c|}
\hline 
Variant & key & nonce & tag & data block & Rounds: $p^{a}$ / $p^{b}$\tabularnewline
\hline 
ASCON-128 & 128 & 128 & 128 & 64 & 12 / 6\tabularnewline
\hline 
ASCON-128a & 128 & 128 & 128 & 128 & 12 / 8\tabularnewline
\hline 
\end{tabular}
\end{table}

ASCON has two different variants for different message lengths: ASCON-128
and ASCON-128a. ASCON-128 uses a message length of 64 bits while ASCON-128a
uses a message length of 128 bits {[}1{]}. The parameter sets for
the different ASCON variants are shown in Table I. As shown, the major
difference is the sizes of the data block and the number of rounds
($p^{b}$). The encryption process is depicted in Fig. 1 and consists
of four stages: 1) Initialization operation, 2) Associated data processing,
3) Plaintext processing, and 4) Finalization operation for tag generation.
These stages are updated using two 320-bit permutations, denoted as
$p^{a}$ and $p^{b}$, where $a$ and $b$ represent the number of
rounds. The permutations are bit-sliced into five 64-bit register
words, forming the 5-bit internal state. In a complete 12-round ASCON,
the permutations iteratively apply an SPN-based round transformation,
which involves adding round constants, applying the substitution layer,
and employing the linear layer for diffusion within the internal state.

The substitution layer incorporates a compact and lightweight 5-bit
S-box as shown in Fig. 2. This S-box is applied in parallel 64 times
to update each bit-slice of the internal state. The 5-bit S-box is
designed using Boolean logic, enabling efficient implementation on
both ASIC and FPGA hardware platforms. The linear layer of ASCON updates
each 64-bit word of the internal state by rotating register words
with different shift values and performing a modulo-2 addition on
the shifted word values.

The decryption process is slightly different than the encryption process.
Instead of plaintext processing, there is a ciphertext processing
function, where the plaintext blocks are computed by XORing the ciphertext
block.

\section{Implementation of the Cryptographic Standard ASCON using ChatGPT}

The task of implementing cryptographic algorithms poses significant
challenges, primarily due to their complex designs and strong security
requirements. The continuous advancements of AI have opened a new
venue for implementing these algorithms, especially through ChatGPT.
This section is divided into two major parts. The first one provides
a detailed methodology for implementing a cryptographic algorithm
through ChatGPT, while the second one focuses on the actual implementation
of ASCON in Python using ChatGPT. 

\subsection{Methodology Followed to Implement a Cryptographic Algorithm using
ChatGPT}

The process, outlined in Fig. 3, is categorized into five stages,
each of which contributes to the successful implementation and testing
of ASCON-128. 
\begin{figure*}
\begin{centering}
\includegraphics[scale=0.8]{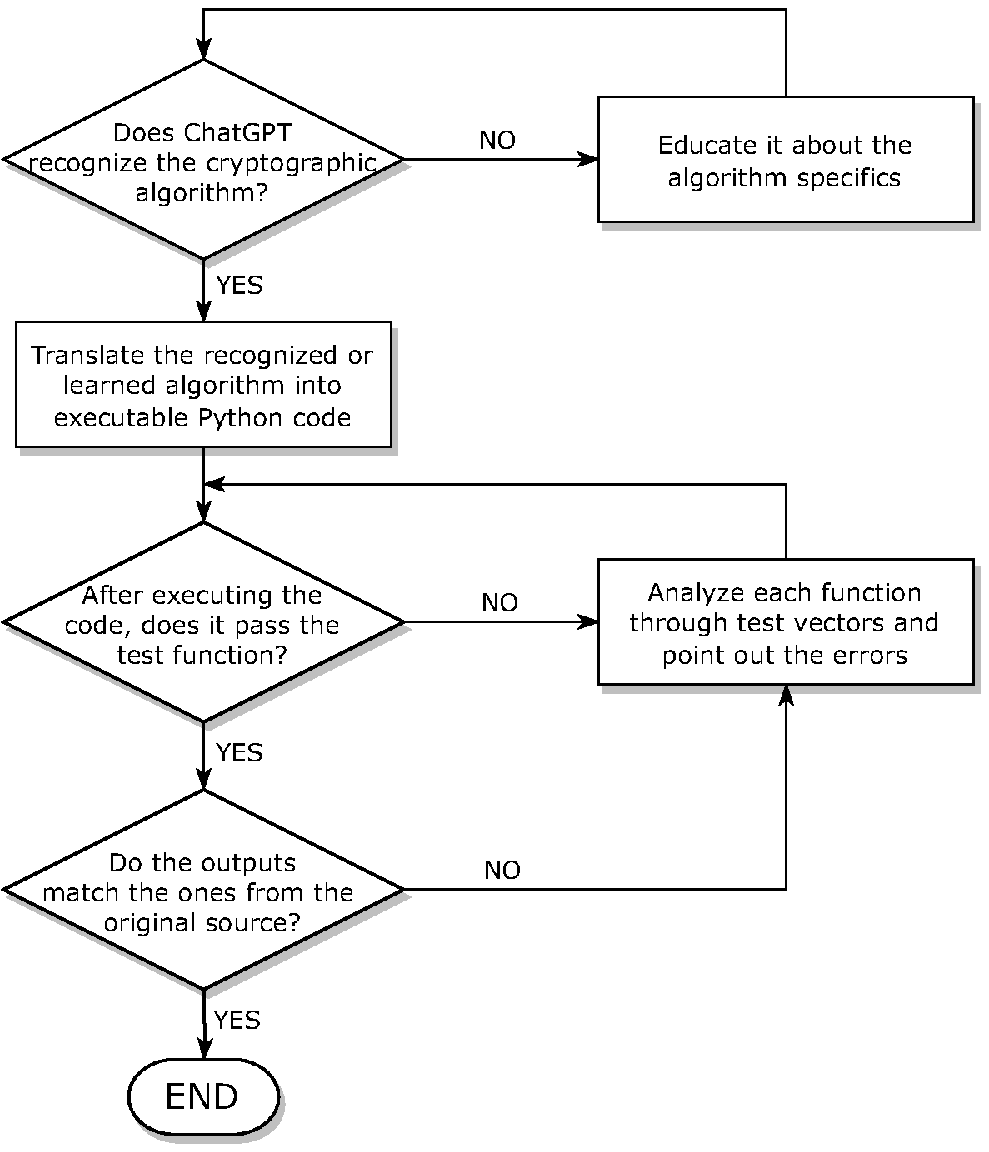}
\par\end{centering}
\caption{Proposed methodology to implement a cryptographic algorithm using
ChatGPT.}
\end{figure*}

\textbf{Stage 1 - Algorithm Recognition:} The initial stage of the
process is to identify the cryptographic algorithm, ASCON, and to
ensure ChatGPT is familiar with it. This step primarily involves the
determination of whether ChatGPT possesses a sufficient understanding
of the algorithm and its main functions and specifications. This understanding
includes a detailed knowledge of the algorithm\textquoteright s specifications
and parameter sets. If ChatGPT does possess a sufficient understanding
of the algorithm, the process advances to Stage 3; if not; it goes
through Stage 2 before moving to the third stage.

\textbf{Stage 2 - Algorithm Education:} If ChatGPT is unfamiliar with
the ASCON algorithm, the second stage involves educating ChatGPT about
the algorithm specifics. Using the capabilities of ChatGPT as a sophisticated
conversational model, the algorithm is taught through comprehensive
textual descriptions, mathematical explanations, and examples. The
detailed specifications of ASCON\textquoteright s functions are communicated
to ChatGPT during this step to ensure a thorough understanding and
prepare it for the following stages. 
\begin{figure*}
\begin{centering}
\includegraphics[scale=0.89]{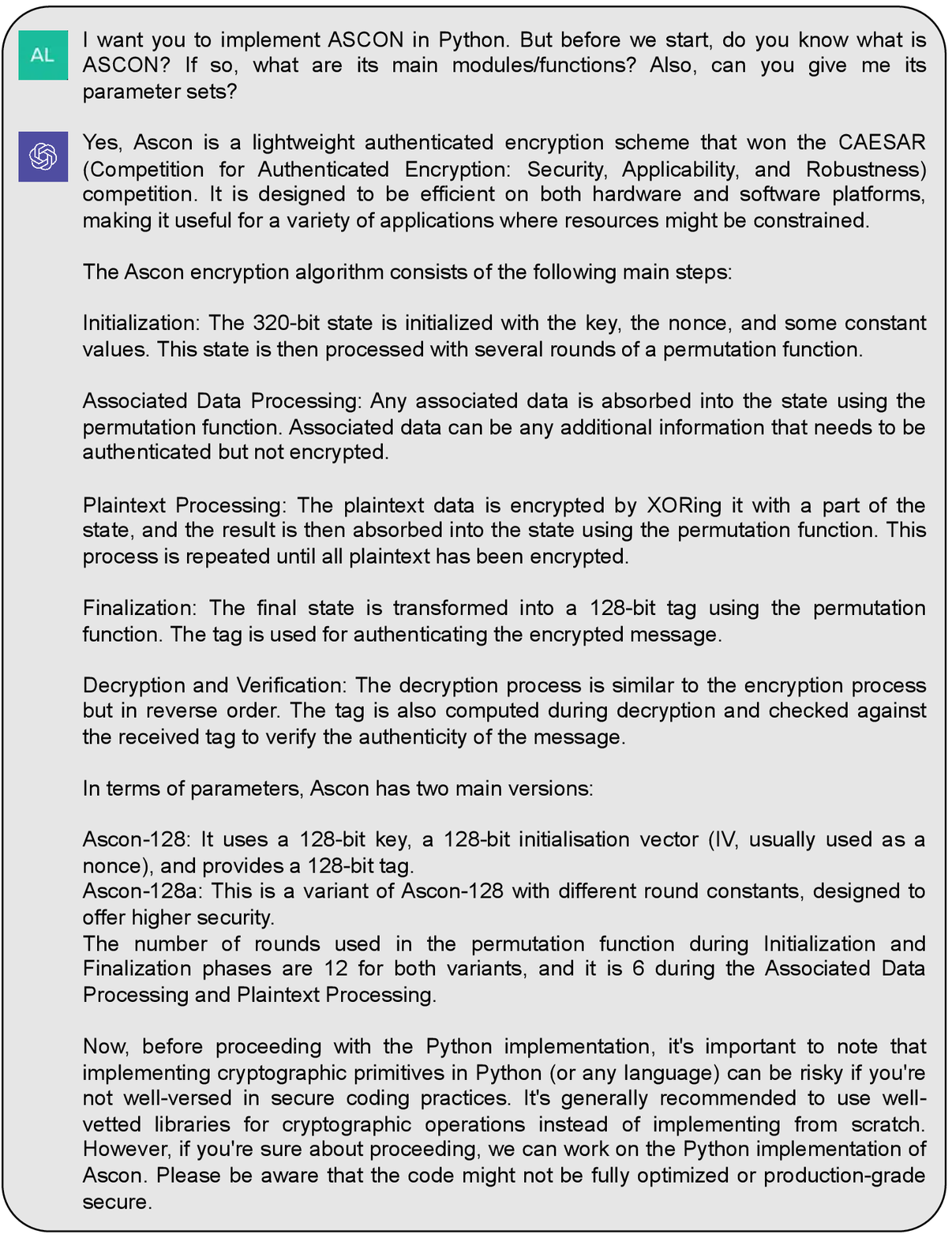}
\par\end{centering}
\caption{ChatGPT\textquoteright s knowledge of ASCON and its main functionalities.}
\end{figure*}

\textbf{Stage 3 - Algorithm Implementation:} In the third stage, we
request ChatGPT to translate the recognized or learned algorithm into
executable Python code. This step consists of asking ChatGPT for the
implementation of ASCON-128 and since it has memory limitations, it
is expected that several prompts may be required for the completion
of this step. Additionally, in this step, a test function from ChatGPT
is requested to validate the functionality of the implemented code.
This allows us to move to the fourth stage: The execution and testing
of the code given by ChatGPT. 

\textbf{Stage 4 - Code Execution and Testing:} The fourth stage involves
executing the Python code and testing its functionality. This two-part
process initially includes basic testing using the test function provided
by ChatGPT. This function needs to compare the original plaintext
with the plaintext derived after decrypting the ciphertext and verifying
the generated tag. Subsequently, the outputs from our Python code
are compared with the results of the original ASCON-128 implementation
provided by its authors. In case of any discrepancies in the output
during either part of the testing, we proceed to the next stage. 
\begin{figure*}
\begin{centering}
\includegraphics[scale=0.89]{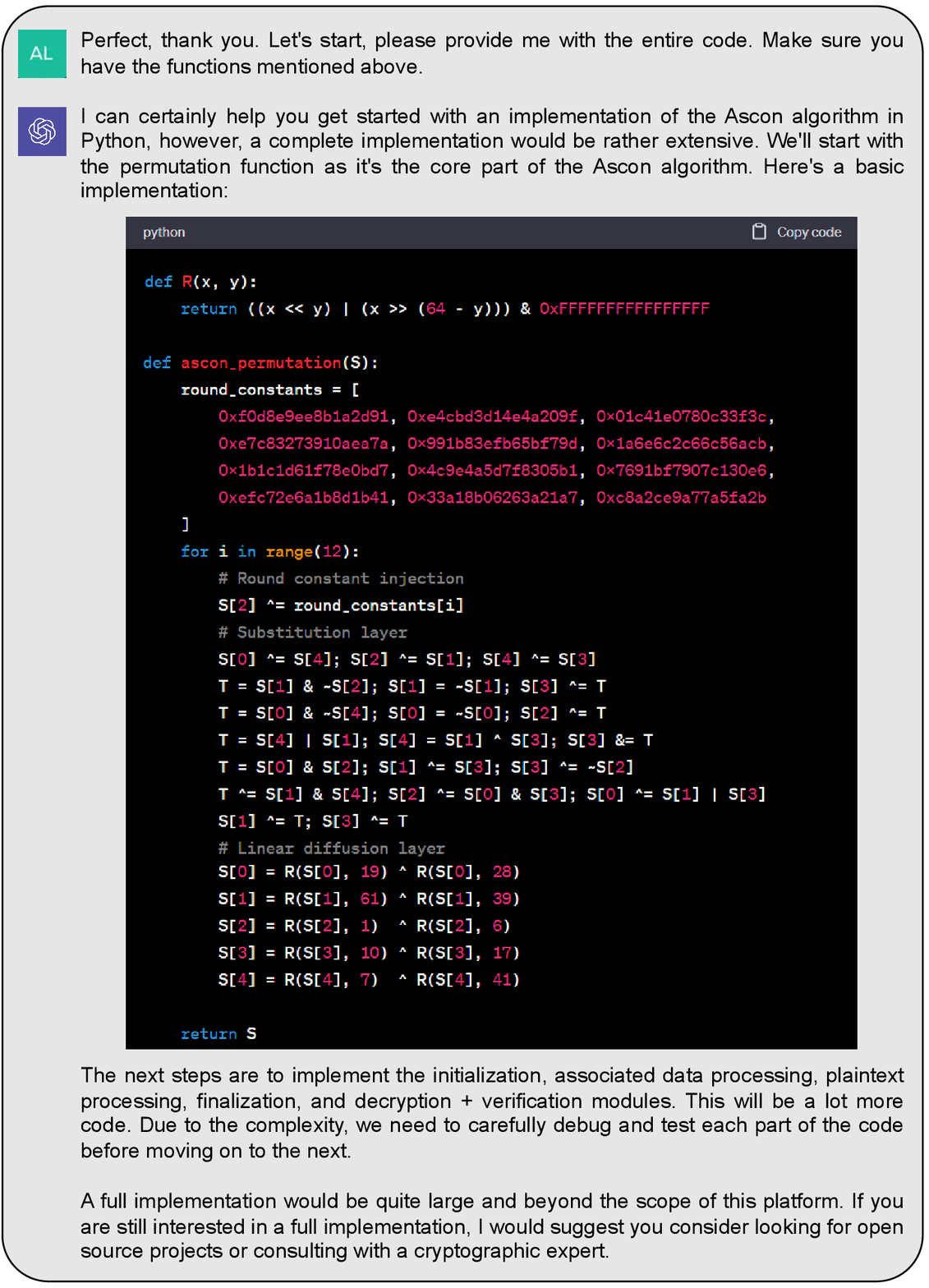}
\par\end{centering}
\caption{The\textit{ R} and \textit{ASCON\_permutation }functions provided
by ChatGPT.}
\end{figure*}

\textbf{Stage 5 - Code Review and Debugging:} If there are any errors
or discrepancies, we examine each of the functions in the context
of ASCON-128\textquoteright s specifications. This involves ensuring
that the code matches the correct specifications of ASCON, as well
as adding test values in different functions to verify their correct
operation. The functions are debugged in the following order: The
initialization function, which should invoke the permutation function;
a process data function, which can operate for both plaintext and
ciphertext or can be divided into two functions; and the finalization
function. Since the process of encryption is carried out first and
then the decryption of the ciphertext takes place, these functions
are revised twice. Also, for simplicity, associated data processing
may be implemented once the version without associated data is operational. 

The steps outlined above englobe the overall process to implement
ASCON-128 in Python using ChatGPT. With a detailed explanation of
each stage, a meticulous approach is adopted to ensure the successful
implementation and verification of ASCON with the assistance of AI.
\begin{figure*}
\begin{centering}
\includegraphics[scale=0.89]{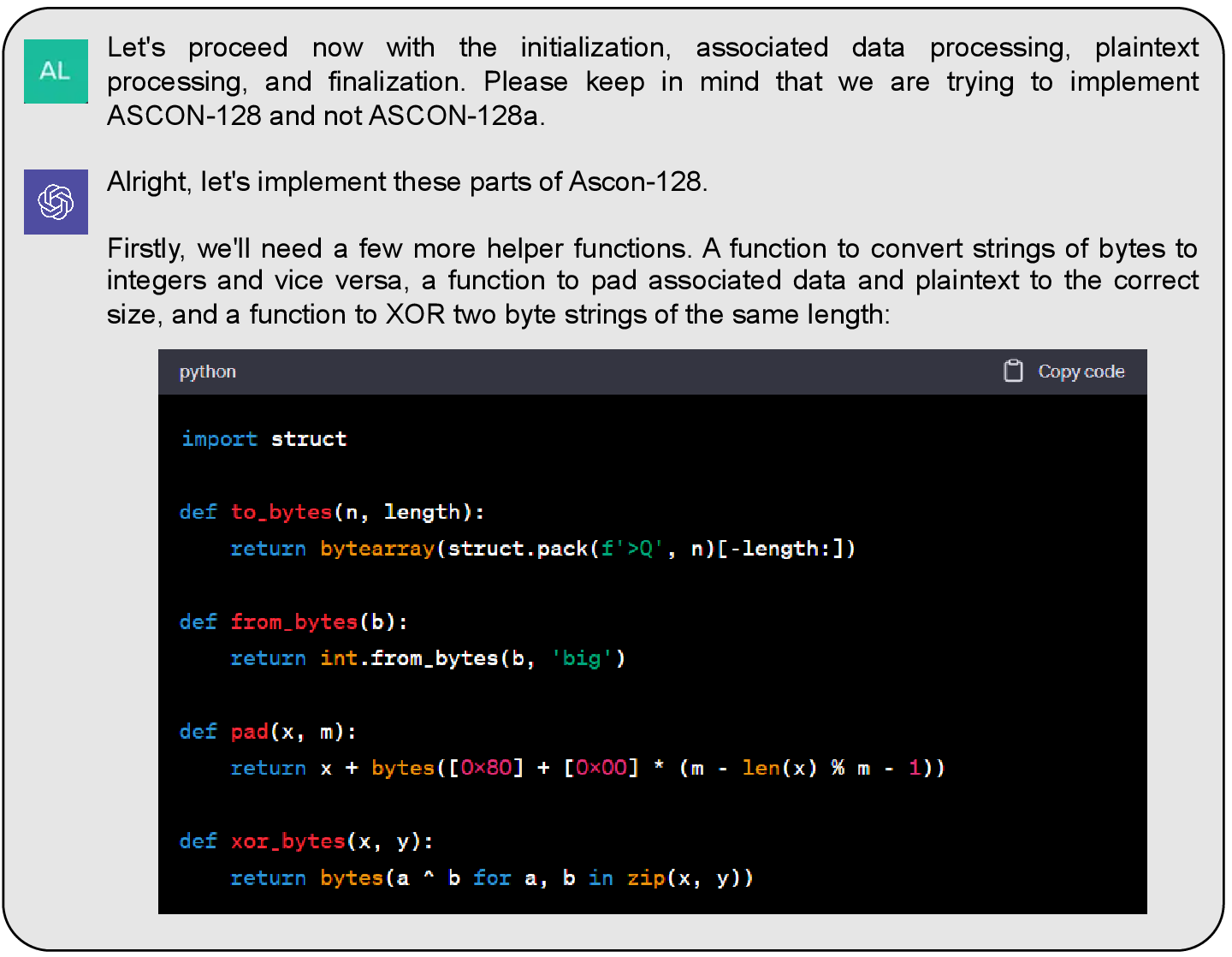}
\par\end{centering}
\caption{Helper functions provided by ChatGPT.}
\end{figure*}

\subsection{Implementation of ASCON in Python Using ChatGPT }

As mentioned above, the first step is querying ChatGPT regarding ASCON
and its specific characteristics. As demonstrated in Fig. 4, ChatGPT
accurately outputs ASCON's algorithm and primary operations, such
as initialization, associated data processing, plaintext processing,
finalization, decryption, and verification, using its knowledge base
up to September 2021. It also recognizes both versions ASCON-128 and
ASCON-128a and their associated parameters. However, it misses mentioning
the number of permutation rounds during the associated data and plaintext
processing phases for ASCON-128a, which performs 8 instead of 6, as
depicted in ASCON-128. Moreover, ChatGPT highlights the potential
risks of implementing cryptographic primitives from scratch, recommending
the use of well-vetted libraries.

Due to ChatGPT's understanding of ASCON, we progress to Stage 3 \textendash{}
the implementation of ASCON-128. When asked to provide the entire
code base with the functions shown in Fig. 4, ChatGPT decides to break
down the code, noting that presenting the complete implementation
would be rather extensive.
\begin{figure*}
\begin{centering}
\includegraphics[scale=0.89]{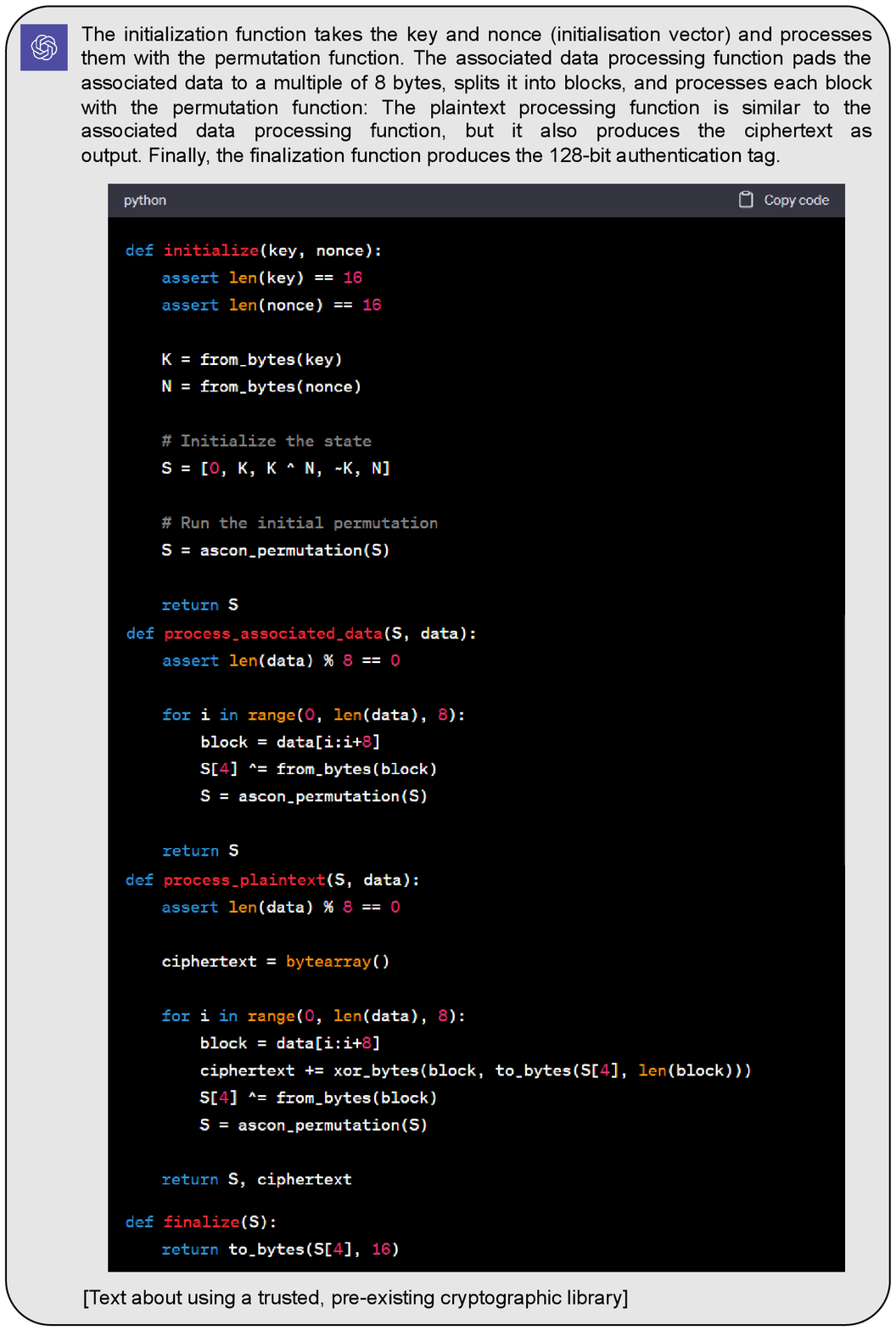}
\par\end{centering}
\caption{The\textit{ initialize, process\_associated\_data, process\_plaintext,}
and \textit{finalize }functions provided by ChatGPT.}
\end{figure*}
\begin{figure*}
\begin{centering}
\includegraphics[scale=0.89]{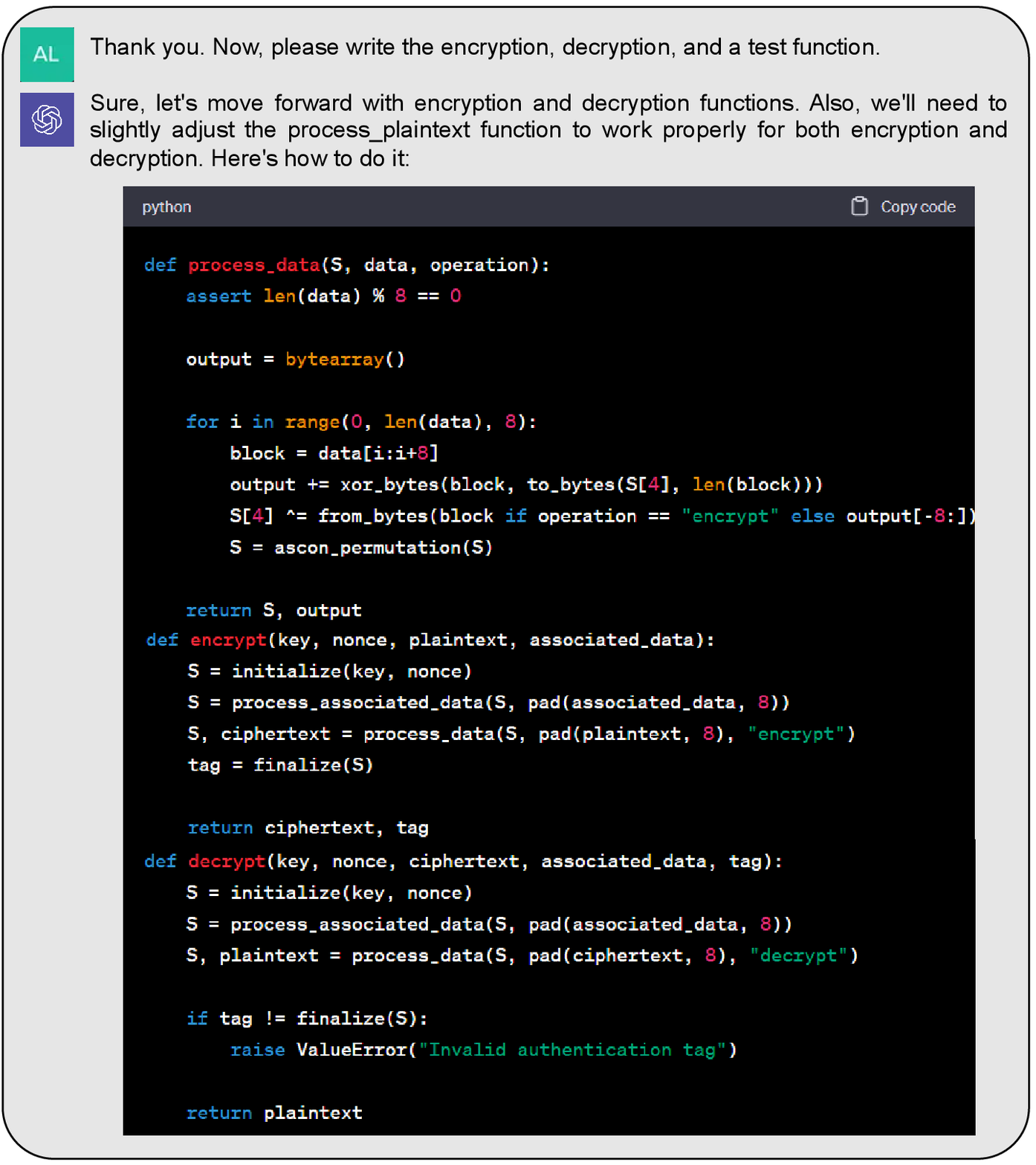}
\par\end{centering}
\caption{The\textit{ process\_data, encrypt,} and \textit{decrypt }functions
provided by ChatGPT.}
\end{figure*}
\begin{figure*}
\begin{centering}
\includegraphics[scale=0.89]{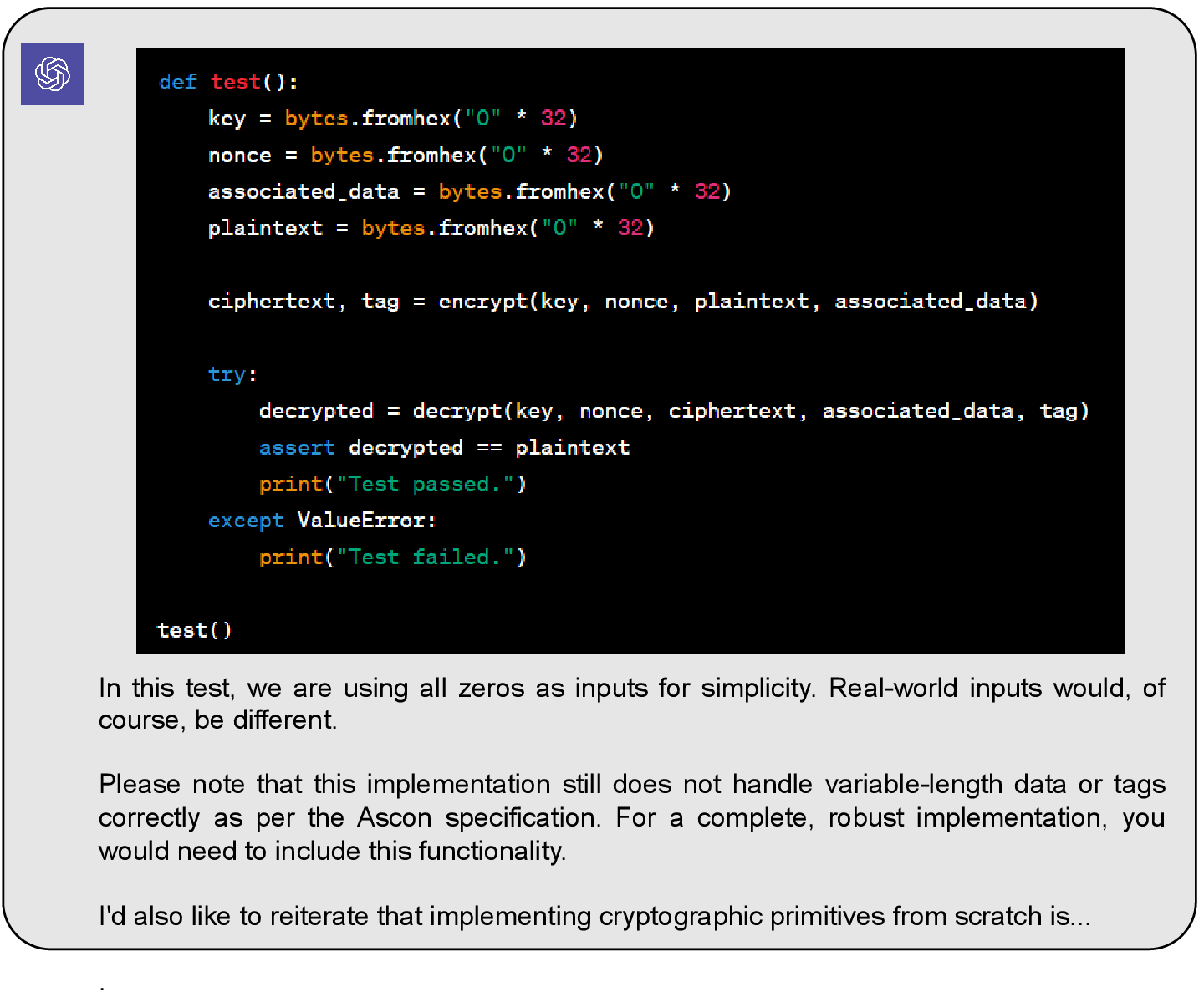}
\par\end{centering}
\caption{The\textit{ test }function provided by ChatGPT.}
\end{figure*}
 As pictured in Fig. 5, it initiates with the \textit{R} function,
which is an abbreviation for rotation and is needed for the linear
diffusion layer of the \textit{ASCON\_permutation} function, also
provided in the same output. The \textit{ASCON\_permutation} function
contains the round constants and the operations executed during the
permutation process: round constant injection, substitution layer,
and linear diffusion layer. Post these implementations, ChatGPT outlines
the other functions needed. Additionally, it recommends again consulting
open-source projects or engaging with a cryptographic expert for guidance.

The next phase requires the implementation of functions to perform
initialization, associated data processing, plaintext processing,
and finalization, pertaining specifically to the ASCON-128 version.
As presented in Fig. 6, before integrating those functions, ChatGPT
suggests and implements the following helper functions: 

- \textit{to\_bytes} function: This function converts strings of bytes
to integers.

- \textit{from\_bytes} function: This function converts integers to
strings of bytes. 

- \textit{pad} function: This function pads associated data and plaintext
to the correct size. 

- \textit{xor\_bytes} function: This function XORs two-byte strings
of the same length.

Upon executing these functions and as shown in Fig. 7, ChatGPT progresses
to implement the \textit{initialize} function, initiating the state
and running the initial permutation process. It follows up with the
\textit{process\_associated\_data} function, processing each 8-byte
block of associated data via the permutation function and the \textit{process\_plaintext}
function that, akin to the previous function, processes each 8-byte
block of data through the permutation function to generate blocks
of ciphertext. The \textit{finalize} function then generates the authentication
tag. 

The concluding functions needed are to encrypt the plaintext, decrypt
the ciphertext, validate the resultant plaintext post-decryption against
the original plaintext, and verify the tag. Since the \textit{process\_plaintext}
function closely resembles the \textit{process\_ciphertext} function,
ChatGPT adapts the \textit{process\_plaintext} into \textit{process\_data}
to account for both encryption and decryption. This is shown in Figs.
8 and 9, along with the implementation of the \textit{encrypt} function,
which invokes the \textit{initialize} function, \textit{process\_associated\_data},
\textit{process\_data}, and \textit{finalize} function to yield the
ciphertext and the tag; the \textit{decrypt} function, which invokes
the same functions as \textit{encrypt} to verify the tag and yield
the plaintext; and the \textit{test} function, which ensures the cryptosystem
functions as expected by invoking the \textit{encrypt} and \textit{decrypt}
functions and using assertions. 

After the provision of the entire code, we advance to Stage 4 \textendash{}
testing. Upon execution, the test fails. Given the multitude of issues
identified in the code, we opt to work first on a simplified version
without associated data and, once the simplified version works as
intended, incorporate the associated data part. As outlined earlier,
each function is tested independently by incorporating print statements
to validate the function outputs. If incorrect, we evaluate if the
code aligns with the ASCON specifications. For the sake of brevity,
some parts will be cropped as many interactions are needed to implement
the entire ASCON cryptosystem. 

The \textit{initialize} function, invoking the \textit{ASCON\_permutation}
function, is the first to be tested. The main issues with the \textit{ASCON\_permutation}
function are: 

- Incorrect constants values.

- Substitution layer not following the specifications. Even after
providing ChatGPT with the specifications, it is still incorrect since
it is not considering that we need to perform bitwise operations in
an unsigned integer context.

- The \textit{R} function, which is invoked in the linear diffusion
layer is performing left rotation instead of right rotation. 

- States are not being XORed in the linear diffusion layer.

- The number of rounds is always 12 but it should be 12 or 6 depending
on the process. 

Moreover, the main issues with the \textit{initialize} function are: 

- Incorrect initial state values and structure. 

- Helper \textit{to\_bytes} function can be modified to always convert
to 8 bytes. 

- When calling the \textit{ASCON\_permutation} function, it should
provide the number of rounds. 

- After the permutation, the internal state is not being XORed with
$0^{*}\:||\:K$. 

The majority of the corrective prompts to accomplish the right implementation
of both \textit{permutation} and \textit{initialize} functions are
depicted in Fig. 10 (located in the Appendix). However, despite the
rectifications, the test still fails, indicating the need to revise
the next major function, the \textit{process\_data} function. 

While we initially focus on the encryption process to validate the
implementation's production of the correct ciphertext, we also identify
the main issues in the \textit{process\_data }function for both encrypting
and decrypting:

- Non-existent padding, which does not allow to accommodate input
data of arbitrary length.

- Incorrect XOR operations. 

- Missing XORing in the decryption process. Moreover, it should perform
an additional XOR operation with $1\:||\:0$ for the last block. 

- Improper trimming of padding. 

- Extra permutation for the last block. 

Fig. 11 (located in Appendix) displays majority of the prompts needed
to correct the \textit{process\_data} function. 

Post validation of the \textit{process\_data} function, we transition
to the \textit{finalize} function. This function has a single line;
therefore, many additions are required: 

- XORing the internal state with $0^{r}\:||\:K\:||\:0^{c}-k$. 

- Executing a 12-round permutation. 

- XORing the permutation result with the key to produce a 16-byte
tag. 

As depicted in Fig. 12 (located in Appendix), most of the prompts
educate ChatGPT about the specifications, rather than indicating the
errors. The \textit{encrypt} and \textit{decrypt} are slightly modified
from the initial version by ChatGPT, taking into account the lack
of associated data. Post validation of the simplified implementation,
the first draft of the functions from Listing 3 only needs to account
for the domain separator $S[4]\:\mathcircumflex=0x1$, which should
be executed after the initialization. The \textit{process\_associated\_data}
function, provided by ChatGPT after the simplified version worked,
only contained a single error: the absence of a permutation after
the final block. The final code can be found at: https://github.com/DrCintas/ASCON-with-ChatGPT.

\section{Conclusion}

To the best of our knowledge this is the first work in the area of
implementing the fairlly recent NIST standard (Feb. 2023) using ChatGPT.
As it has been shown throughout this paper, the results prove that
the implementation of a NIST standard lightweight cryptographic algorithm
using its specifications and ChatGPT is feasible. By using the methodology
proposed in this paper, any individual has the ability to implement
cryptographic algorithms using several prompts in ChatGPT and by having
a deep understanding of the specifications. There is no need of explaining
the fundamentals of ASCON to ChatGPT as it recognizes the cryptographic
algorithm and its main functions. However, prompting deeper mathematical
specifications are needed as the first code provided does not provide
the intended results. Although it is feasible not to prompt any code,
as we have done, sometimes it may be easier to specify ChatGPT the
exact line of the code that is wrong instead of specifying the details
of the entire function, since ChatGPT can modify other parts that
are already correct.

The resulting code from the ChatGPT interactions does not have any
errors and it is able to match all the test vectors when compared
to the original implementation. However, this does not mean that one
should implement their own cryptosystem and rely on ChatGPT to do
so. Furthermore, ChatGPT reminds the users of this and recommends
them to consult open-source projects or engage with a cryptographic
expert for guidance.

\section*{Acknowledgements}

This work was supported by the U.S. federal agency award 60NANB20D013
granted from National Institute of Standards and Technology (NIST).

\appendix{The following figures contain the prompts used to correct the implemetation
of ASCON in Python using ChatGPT.
\begin{figure*}
\begin{centering}
\includegraphics[scale=0.9]{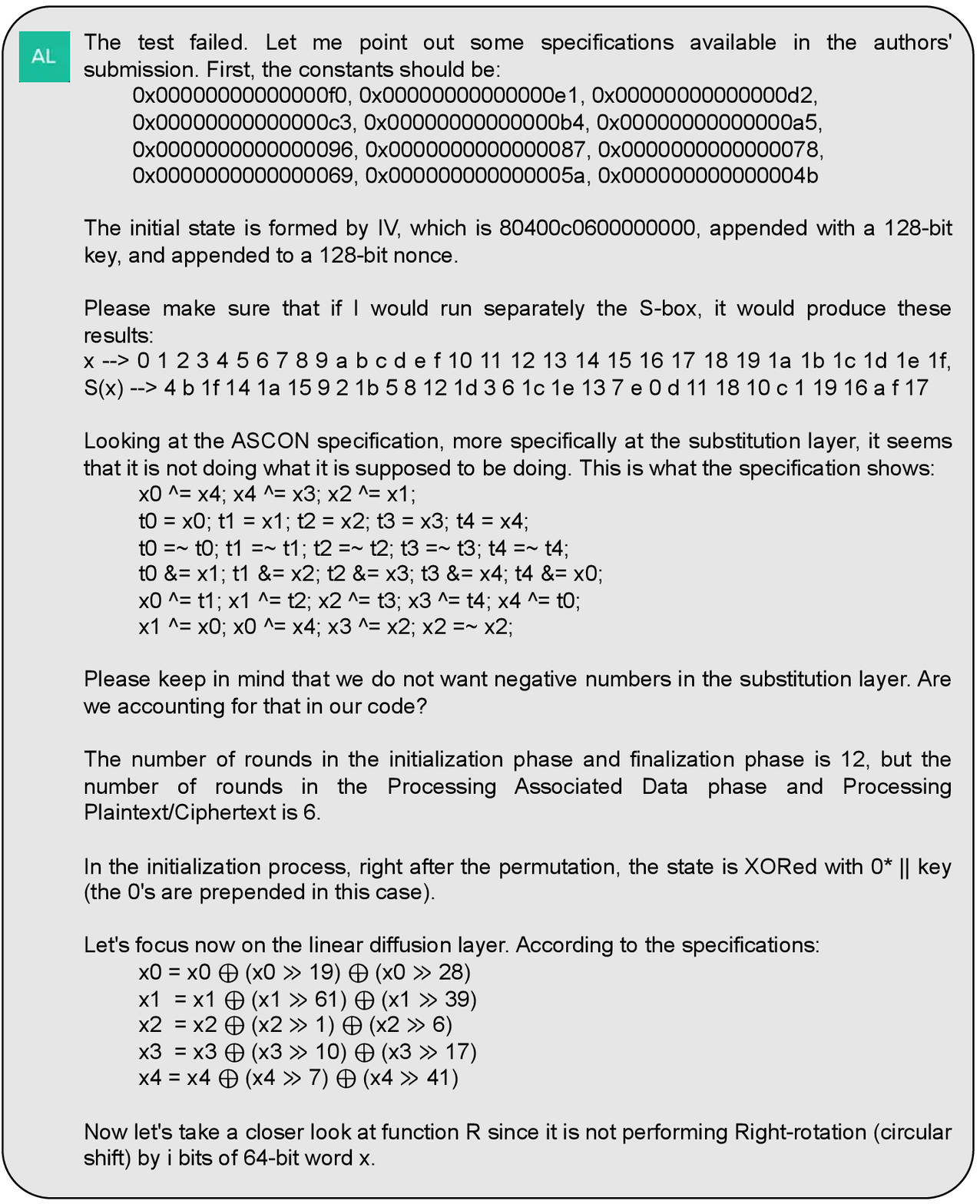}
\par\end{centering}
\caption{Prompts provided to correct the \textit{initialize, ASCON\_permutation,}
and \textit{R} functions.}
\end{figure*}
\begin{figure*}
\begin{centering}
\includegraphics[scale=0.9]{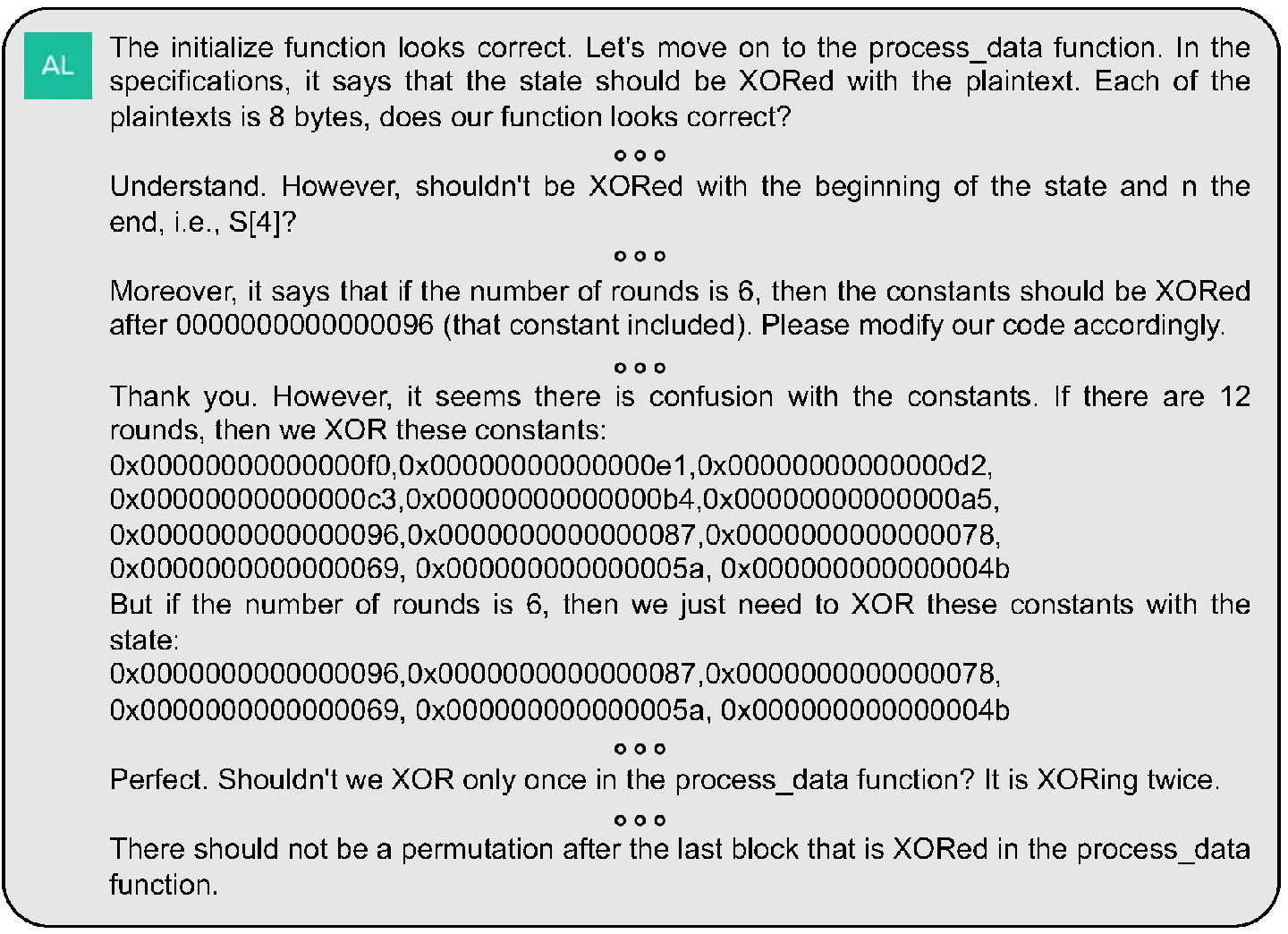}
\par\end{centering}
\caption{Prompts provided to correct the \textit{process\_data} function.}
\end{figure*}
\begin{figure*}
\begin{centering}
\includegraphics[scale=0.9]{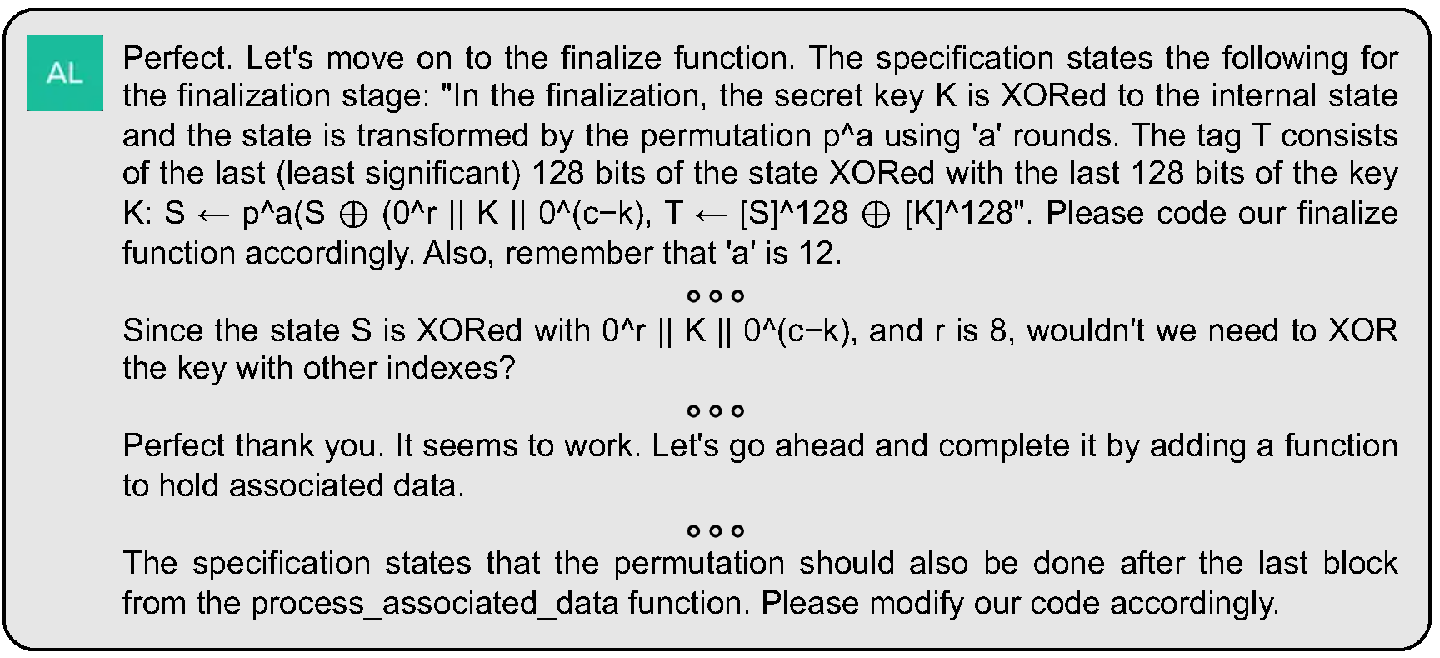}
\par\end{centering}
\caption{Prompts provided to correct the \textit{finalize} function.}
\end{figure*}
}
\end{document}